\documentclass[aps,pre,twocolumn,floatfix,showpacs]{revtex4}
\usepackage{graphicx}

\begin{document}

\title{The Domain Chaos Puzzle and the Calculation of the Structure Factor and Its Half-Width}
\author{Nathan Becker}
\email{nbecker@physics.ucsb.edu}
\address{Department of Physics and iQCD, University of California, Santa
Barbara, California 93106}
\author{Guenter Ahlers}
\email{guenter@physics.ucsb.edu}
\address{Department of Physics and iQCD, University of California, Santa
Barbara, California 93106}

\begin{abstract}
The disagreement of the scaling of the correlation length $\xi$ between experiment and the Ginzburg-Landau (GL) model for domain chaos was resolved.
The Swift-Hohenberg (SH) domain-chaos model was integrated numerically to acquire test images to study the effect of a finite image-size on the
extraction of $\xi$ from the structure factor (SF).  The finite image size had a significant effect on the SF determined with the 
Fourier-transform (FT) method.  The maximum entropy method (MEM) was able to overcome this finite image-size
problem and produced fairly accurate SFs for the relatively small image sizes provided by experiments.

Correlation lengths often have been determined from the second moment of the SF of chaotic patterns because the functional form of the SF is not known. Integration of several test functions provided
analytic results indicating
that this may not be a reliable method of extracting $\xi$. For both a Gaussian and a squared SH form, the correlation length $\overline{\xi} \equiv 1/\sigma$, determined from the variance $\sigma^2$ of the SF, 
has the same dependence on the control parameter $\varepsilon$
as the length $\xi$ contained explicitly in the functional forms. However, for the SH and the Lorentzian forms we find $\overline{\xi} \sim \xi^{1/2}$. 

Results
for $\xi$ determined from new experimental data by fitting the functional forms directly to the
experimental SF yielded $\xi\sim\varepsilon^{-\nu}$ with $\nu\simeq1/4$ for all four functions in the case of
the FT method, but $\nu\simeq1/2$, in agreement with the GL prediction, in the the case of the MEM.
Over a wide range of $\varepsilon$ and wave number $k$, the experimental
SFs collapsed onto a unique curve when appropriately scaled by $\xi$.
\end{abstract}
\maketitle

\section{Introduction}
Spatially extended nonlinear non-equilibrium systems continue to be
of great interest because they yield qualitatively different phenomena
that do not occur in linear systems \cite{CH93}. One of these phenomena
is spatio-temporal chaos. The examples of spatio-temporal chaos found
in Rayleigh-B\'{e}nard convection (RBC) lend themselves to particularly
detailed experimental study under exceptionally well controlled external
conditions \cite{BPA00}. RBC occurs in a thin horizontal layer of
fluid with thickness $d$ heated from below when the temperature difference
$\Delta T$ exceeds a critical value $\Delta T_{c}$ \cite{SC}.

\begin{figure}[h!]
\begin{center}\includegraphics[width=3.29in,keepaspectratio]{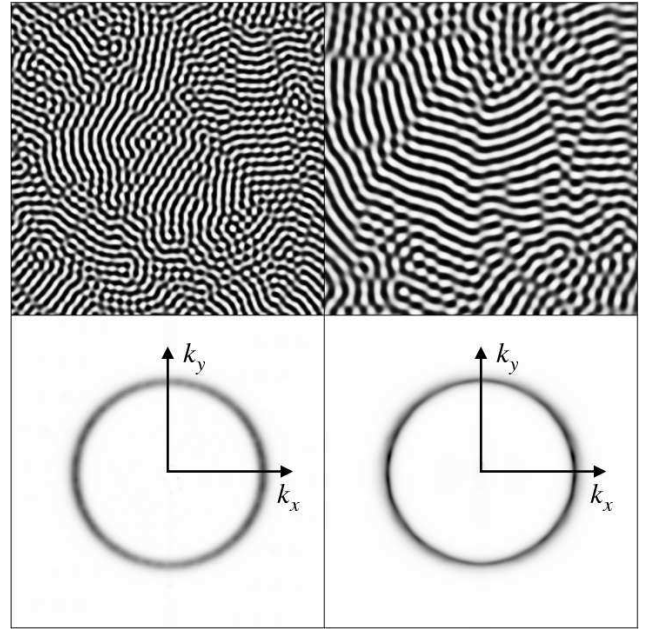}\end{center}
\caption{\label{cap:Pattern}All images are for $\Omega=17.7$ and $\varepsilon=0.05$. Upper left: shadowgraph image of size $60d\times60d$
from $\Gamma=61.5$. A movie is available \cite{MOV1}. Upper right: shadowgraph image of
size $44d\times44d$ from $\Gamma=36$.
Lower left: the average SF $S\left(k_x,k_y\right)$
computed using the FT method and a Kaiser-Bessel window with $\alpha=2.5$ and averaged over 4096
images like the one in the upper left.  Lower right: the average SF from same data as shown
in lower left, except $S\left(k_x,k_y\right)$ was computed with the MEM.}
\end{figure}

Particularly noteworthy is the state that occurs in RBC when the fluid
layer of density $\rho$ and shear viscosity $\eta$ is rotated about
a vertical axis with angular frequency $\omega$. When the dimensionless
frequency $\Omega\equiv\omega d^{2}\rho/\eta$ exceeds a critical
value, then the pattern immediately above onset consists of disordered
domains of convection rolls known as domain chaos. This is illustrated
in Fig.~\ref{cap:Pattern}a and \ref{cap:Pattern}b. Within each domain the roll orientation
is more or less uniform; but different domains have different orientations
\cite{KL1,KL2,CB1,HB,BH,HEA1,HEA2,HPAE98}. The domains are unstable and undergo a persistent
but irregular dynamics. In this case, the time-averaged mean-square
velocities and temperature deviations from the conduction state grow
continuously from zero as $\Delta T_{c}$ is exceeded, i.e. the bifurcation
is supercritical.

Chaos immediately above a supercritical bifurcation offers a unique
opportunity for theoretical study because weakly-nonlinear theories
are expected to be applicable. These theories, in the form of Ginzburg-Landau
(GL) or Swift-Hohenberg (SH) equations, by virtue of their general
structure and from their numerical solutions, predict that the inverse
half-width at half-height $\xi$ of the structure factor (SF, the power spectrum of the pattern) should
vary as $\varepsilon^{-\nu}$ with $\nu=1/2$ as $\varepsilon\equiv\Delta T/\Delta T_{c}-1$
vanishes \cite{Friedrich,Fantz,Neufeld,Cross,CMT,PPS97,PPS972}, and that it is the only length
scale in the problem. Thus it was particularly disappointing that
measurements for domain chaos disagreed with this expectation
\cite{HEA1}. Determinations of a correlation length $\bar{\xi}$ based
on the variance of the SF found $\bar{\xi}\propto\varepsilon^{-\nu_{eff}}$
with $\nu_{eff}\simeq1/4$ rather than 1/2.

Several possible explanations of the apparent disagreement with theory
were explored by various authors. Hu et al. proposed that defects
and fronts injected by the side wall into the bulk may play an important
role \cite{HPAE98}. Laveder et al. demonstrated that additive white
noise intended to mimic the effect of wall defects decreases $\nu$
to the value of $\nu_{eff}$ measured experimentally provided that
the noise level is sufficiently large \cite{LPPS99}. Recent unpublished
experiments in our group using a sample with a radially ramped spacing
\cite{BMCA99,BMOA05} render the side-wall-injected defect idea an unlikely
explanation. On the basis of numerical simulation using a SH equation
Cross et al. proposed that the finite size of the experimental convection
sample decreases the effective value of $\nu$ \cite{CLM01}.

Experimental determinations of a characteristic length scale $\overline{\xi}$
usually are based on numerical estimates of the variance $\sigma^{2}$
of the SF, with $\overline{\xi}=1/\sigma$ \cite{HEA1,MBCA93}. This is so because the precise analytic form of $S(k)$ is not known in the nonlinear regime above onset. 
In Sect.~\ref{sub:Analytic_Moments} we examine the relationship between $\overline{\xi}$ and the
length $\xi$ that appears explicitly in various functional forms that might be used as approximations to the SF. We find that $\overline{\xi} \sim \xi^{1/2}$ when $\xi$ is the length appearing in the SF of the linear GL or SH equation. Thus, in those cases $\overline{\xi}$ 
does not have the same $\varepsilon$-dependence as $\xi$. However, we find that  $\overline{\xi} \sim \xi$ for the square of the linear SH SF and for a Gaussian form for the SF. The results suggest that the success of the moment method of data analysis depends upon the rate at which the SF decreases at large $k$.

In Sect.~\ref{sub:SkMethods} we discuss two methods of determining the SF from patterns.  The standard method,
using the Fourier power spectrum, is sensitive to the limited size of the experimental images \cite{FNsize}.  To overcome
this limitation we also employed the maximum entropy method (MEM), which is discussed in detail in that section.  The
benefits of the MEM are exhibited through analysis of images from a SH simulation detailed in Sect.~\ref{sub:SimSH}.
We find that the MEM is quite powerful in its ability to overcome the finite image-size problem.

In Sect.~\ref{sub:Experimental_Measurement_xi}
we present new experimental results for patterns in the domain-chaos state.  We compare results from the Fourier analysis to the results from the MEM. We determined $\xi$ by fitting several possible functional
forms for the SF to the data and found that the results for $\xi$ are not very sensitive to the form of the fitting function. In the case of the Fourier analysis, the fits did not change the result \cite{HEA1} $\nu\simeq1/4$ that
 had been found before by the moment method. The reason why the moment method also gave this result can be found in the behavior of $S(k)$ at large $k$, where $S(k)$ 
drops off fast enough for $\overline \xi$ to be essentially proportional to $\xi$.  In the case of the MEM, we found
that $\nu\simeq1/2$ as expected.  This indicates that the findings from Fourier analysis are dominated by the
finite image-size effect.
 
 We also examined the maximum height $B$ of the SF and found $B \sim \varepsilon^\beta$ with 
 $\beta \simeq 3/4$ in the case of Fourier analysis. The two results $\nu \simeq 1/4$ and $\beta \simeq 3/4$ conspire to retain the 
 expected dependence of the total power on $\varepsilon^{\nu+\beta}$ with $\nu+\beta = 1$ even though both $\nu$ and $\beta$ do not have the expected value 1/2.  Once again, we find that the MEM overcomes the finite data length yielding 
$\beta\simeq1/2$.  We also find that the result for $\nu$ and $\beta$ are not strongly dependent on $\Omega$ for
either the Fourier analysis or the MEM.  
Thus we conclude that, in light of careful analysis using the MEM, the experimental domain-chaos state does possess a length scale in agreement with
the prediction from the GL model.

Finally, in Sect.~\ref{sub:Scaling_Collapse}, we examine the extent to which the SF can be represented by a unique scaling function over a range of $\varepsilon$ and $k$. We obtain excellent collapse of the data, both in the case of Fourier analysis and the MEM, when the results for $\xi$ obtained from a fit to the SF at each $\varepsilon$ are used to scale the SF. 

\section{\label{sub:Analytic_Moments}Moments of Analytic Functions}

In order to better understand the consequences of using numerical
moments to compute the correlation length, we derived analytic expressions for the zeroth,
first, and second moments of several proposed forms of the SF. This
yielded results for the correlation length $\overline{\xi}$
based on moments that could be compared with the correlation
length $\xi$ that occurs directly in the functional form chosen for the SF.
We found that the $\overline{\xi}\sim\xi^{m}$ where $m$ depended
on the particular form of the structure factor. This suggests that
using moments to measure the correlation length does not necessarily
yield $\overline{\xi}$ with the $\varepsilon$ dependence
implied by the GL model.

We investigated four particular forms of the SF. The exact form for domain chaos is not known. For our purposes one useful form is the SF of the linear SH equation

\begin{equation}
S\left(k\right)=\frac{4k_{0}^{2}B}{\xi^{2}\left(k^{2}-k_{0}^{2}\right)^{2}+4k_{0}^{2}}\textrm{.}
\label{eq:SHSF}
\end{equation}

\noindent It is an excellent approximation to the SF of the linearized full
equations of motion (the Boussinesq equations) in the presence of
additive noise for RBC below onset \cite{OS02}. Another useful form
is the squared SH SF

\begin{equation}
S\left(k\right)=\left[\frac{4k_{0}^{2}\sqrt{B}}{\xi_{s}^{2}\left(k^{2}-k_{0}^{2}\right)^{2}+4k_{0}^{2}}\right]^{2}
\label{eq:squaredSHSF}
\end{equation}

\noindent used by some authors for fits to numerical
results in order to estimate a half width at half height $\delta k$ of the distribution
\cite{sqSH,sqSH2}. In the limit of large $\xi$, the correlation length
$\xi$ in Eq.~\ref{eq:SHSF} approaches $1/\delta k$, but $\xi_{s}$
in Eq.~\ref{eq:squaredSHSF} approaches $\sqrt{\sqrt{2}-1}/\delta k$
for large $\xi_{s}$. When we directly compare $\xi$ and $\xi_{s}$, as in Fig.~\ref{cap:xi_fitting} below, 
we divide $\xi_{s}$ by $\sqrt{\sqrt{2}-1}$ so that it approaches
$1/\delta k$. 

We also considered a Gaussian form

\begin{equation}
S\left(k\right)=B\exp\left[-\left(\ln2\right)\left(k-k_{0}\right)^{2}\xi^{2}\right]\label{eq:GaussianSF}\end{equation}

\noindent and a Lorentzian form

\begin{equation}
S\left(k\right)=\frac{B}{\xi^{2}\left(k-k_{0}\right)^{2}+1}\textrm{.}\label{eq:LorentzianSF}\end{equation}

\noindent The latter is the SF of the one-dimensional linearized GL equation. The correlation length $\xi$ of both the Gaussian and the Lorentzian
is exactly equal to $1/\delta k$ regardless of the size of $\xi$.
For all four of these functions we note that the position of the peak
of $S\left(k\right)$ is at $k_{0}$ and $S\left(k_{0}\right)=B$.

The total power is $P=2\pi\int_{0}^{\infty}S\left(k\right)kdk$. In
the case of the SH form,

\begin{equation}
P=\frac{2\pi k_{0}B}{\xi}\left[\frac{\pi}{2}-\tan^{-1}\left(-\frac{\xi k_{0}}{2}\right)\right]\label{eq:SH_exactpower}\end{equation}

\noindent which reduces to\begin{equation}
P=\frac{2\pi^{2}k_{0}B}{\xi}\left[1-\frac{2}{\pi\xi k_{0}}+\mathcal{{O}}\left(\frac{1}{\xi^{3}}\right)\right]\label{eq:SH_expandedpower}
\end{equation}

\noindent in the limit of large $\xi$. In the case of the squared SH form,

\begin{equation}
P=2\pi k_{0}^{2}B\left\{ \frac{1}{4+\xi_{s}^{2}k_{0}^{2}}+\frac{1}{2\xi_{s}k_{0}}\left[\frac{\pi}{2}-\tan^{-1}\left(-\frac{\xi_{s}k_{0}}{2}\right)\right]\right\} \label{eq:squaredSH_exactpower}\end{equation}

\noindent which reduces to

\begin{equation}
P=\frac{\pi^{2}k_{0}B}{\xi_{s}}\left[1+\mathcal{{O}}\left(\frac{1}{\xi_{s}^{3}}\right)\right]\label{eq:squaredSH_expandedpower}\end{equation}

\noindent in the limit of large $\xi_{s}$. 

The first moment is $\overline{k}=(2\pi/P)\int_{0}^{\infty}S\left(k\right)k^{2}dk$.

\noindent In the case of the SH form

\begin{equation}
\overline{k}=k_{0}\frac{\frac{2}{\xi k_{0}}\sin\left[\frac{1}{2}\tan^{-1}\left(\frac{2}{\xi k_{0}}\right)\right]+\cos\left[\frac{1}{2}\tan^{-1}\left(\frac{2}{\xi k_{0}}\right)\right]}{\left(1+\frac{4}{\xi^{2}k_{0}^{2}}\right)^{1/4}\left[\frac{1}{2}-\frac{1}{\pi}\tan^{-1}\left(-\frac{\xi k_{0}}{2}\right)\right]}\label{eq:SH_exactkbar}\end{equation}

\noindent which reduces to

\begin{equation}
\overline{k}=k_{0}\left[1+\frac{2}{\pi\xi k_{0}}+\mathcal{{O}}\left(\frac{1}{\xi^{2}}\right)\right]\label{eq:SH_expandedkbar}\end{equation}

\noindent in the limit of large $\xi$.

In the case of the squared SH form

\begin{widetext}

\begin{equation}
\overline{k}=k_{0}\frac{\pi\sin\left[\frac{1}{2}\tan^{-1}\left(\frac{2}{\xi_{s}k_{0}}\right)\right]+\pi\xi_{s}k_{0}\cos\left[\frac{1}{2}\tan^{-1}\left(\frac{2}{\xi_{s}k_{0}}\right)\right]}{2\xi_{s}^{2}k_{0}^{2}\left(1+\frac{4}{\xi_{s}^{2}k_{0}^{2}}\right)^{1/4}\left(\frac{1}{4+\xi_{s}^{2}k_{0}^{2}}+\frac{1}{2\xi_{s}k_{0}}\left[\frac{\pi}{2}-\tan^{-1}\left(-\frac{\xi_{s}k_{0}}{2}\right)\right]\right)}\label{eq:squaredSH_exactkbar}\end{equation}

\end{widetext}

\noindent which reduces to

\begin{equation}
\overline{k}=k_{0}\left[1-\frac{1}{2\xi_{s}^{2}k_{0}^{2}}+\mathcal{{O}}\left(\frac{1}{\xi_{s}^{3}}\right)\right]\label{eq:squaredSH_expandedkbar}\end{equation}

\noindent in the limit of large $\xi_{s}$.

The second moment does not converge for the SH form but it does for
the squared SH form. In the case of squared SH, $\sigma^{2}\equiv\overline{k^{2}}-\overline{k}^{2}$,
where

\begin{equation}
\overline{k^{2}}=k_{0}^{2}\frac{\frac{1}{\xi_{s}^{2}k_{0}^{2}}+\frac{1}{2\xi_{s}k_{0}}\left[\frac{\pi}{2}-\tan^{-1}\left(-\frac{\xi_{s}k_{0}}{2}\right)\right]}{\frac{1}{4+\xi_{s}^{2}k_{0}^{2}}+\frac{1}{2\xi_{s}k_{0}}\left[\frac{\pi}{2}-\tan^{-1}\left(-\frac{\xi_{s}k_{0}}{2}\right)\right]}\label{eq:squaredSH_exactk2bar}\end{equation}

\noindent and Eq.~\ref{eq:squaredSH_exactkbar} gives $\overline{k}^{2}$. For
large $\xi_{s}$,

\begin{equation}
\sigma^{2}=\frac{1}{\xi_{s}^{2}}\left[1-\frac{8}{3\pi\xi_{s}k_{0}}+\mathcal{{O}}\left(\frac{1}{\xi_{s}^{2}}\right)\right]\label{eq:squaredSH_expandedsigma2}\end{equation}

\noindent so that $\overline{\xi}_{s}\sim\xi_{s}$.

In order to compute a similar expression for the SH form,
we introduced a cutoff $k_{C}$ so that the second moment
remained finite. In that case $\sigma^{2}=2\pi P^{-1}\int_{0}^{k_{C}}S\left(k\right)k^{3}dk -\overline{k}^{2}$
so that

\begin{widetext}

\begin{equation}
\sigma^{2}=\frac{k_{0}^{2}\left\{ \tan^{-1}\left(\frac{\xi}{2k_{0}}\left[k_{C}^{2}-k_{0}^{2}\right]\right)+\tan^{-1}\left(\frac{\xi k_{0}}{2}\right)\right\} +\frac{k_{0}}{\xi}\ln\left(\frac{4k_{0}^{2}+\xi^{2}\left[k_{C}^{2}-k_{0}^{2}\right]^{2}}{4k_{0}^{2}+\xi^{2}k_{0}^{4}}\right)}{\tan^{-1}\left(\frac{\xi k_{C}^{2}}{2k_{0}}-\frac{\xi k_{0}}{2}\right)-\tan^{-1}\left(-\frac{\xi k_{0}}{2}\right)}-\overline{k}^{2}\label{eq:SH_exactsigma2}\end{equation}

\end{widetext}

\noindent and in the limit of large $\xi$

\begin{equation}
\sigma^{2}\simeq\frac{1}{\xi}\left[\frac{k_{0}}{\pi}\ln\left(\frac{\left[k_{C}^{2}-k_{0}^{2}\right]^{2}}{k_{0}^{4}}\right)-\frac{4k_{0}}{\pi}\right]+\mathcal{{O}}\left(\frac{1}{\xi^{2}}\right)\textrm{.}\label{eq:SH_approxsigma2}\end{equation}

\noindent Equation \ref{eq:SH_approxsigma2} indicates that in the case of the SH form $\bar{\xi}\sim\sqrt{\xi}$
regardless of the cutoff $k_C$. Although we do not present the details
here, we also found that $\overline{\xi}\sim\sqrt{\xi}$ for the Lorentzian
form but $\overline{\xi}\sim\xi$ for the Gaussian form.

\begin{figure}
\begin{center}\includegraphics[width=3.33in,keepaspectratio]{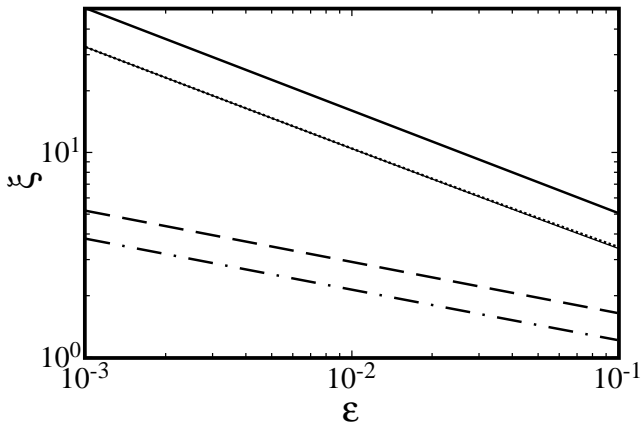}\end{center}
\caption{\label{cap:analytic_plots}$\overline{\xi}$ computed from integration
of the SH SF and the squared SH SF. The analytic results were
evaluated with $\xi=1.6\varepsilon^{-1/2}$, $k=\pi$, and $k_{C}=14$.
Solid line: $\xi=1.6\varepsilon^{-1/2}$ (chosen to match the solid 
line in Fig.~\ref{cap:xi_fitting}). Dashed line: $\overline{\xi}$
from SH computed with Eq.~\ref{eq:SH_exactsigma2} using $k_{C}=14$
and computing $\overline{k}$ numerically so that the integration
is done only over the range $0<k<14$ for consistency. Dashed-dotted
line: Approximate $\overline{\xi}$ from SH computed with Eq.~\ref{eq:SH_approxsigma2}.
Dotted line: $\overline{\xi}_{s}$ computed from Eq.~\ref{eq:squaredSH_exactkbar}
and Eq.~\ref{eq:squaredSH_exactk2bar}. Thin solid line that passes
almost directly through dotted line: Approximate $\overline{\xi}_{s}$
from squared SH computed from Eq.~\ref{eq:squaredSH_expandedsigma2}. }
\end{figure}

Figure \ref{cap:analytic_plots} compares these results for $\overline{\xi}$
computed by numerically integrating the SH SF and the squared SH SF.
The lowest order expansion for the squared SH SF in Eq.~\ref{eq:squaredSH_expandedsigma2}
is an excellent approximation to the exact result as it lands almost
directly on top of it. However, although the exponent is $-1/2$,
it does give a $\overline{\xi}_{s}$ that is smaller than the input
$\xi$ (shown as the solid line in Fig.~\ref{cap:analytic_plots})
because of the extra constant $\sqrt{\sqrt{2}-1}$ discussed earlier.
The approximate result for the SH SF (from Eq.~\ref{eq:SH_approxsigma2})
is not very close to the exact result (from Eq.~\ref{eq:SH_exactsigma2})
because the exact result for $\overline{k}$ (Eq.~\ref{eq:SH_exactkbar})
was used to compute Eq.~\ref{eq:SH_approxsigma2} while the dashed
curve shown in Fig.~\ref{cap:analytic_plots} was computed with a
finite cutoff for $\overline{k}$ to match the cutoff used in the
integral for $\overline{k^{2}}$ for the sake of consistency. Nevertheless,
the approximate result is parallel to the exact result with cutoff
indicating that regardless of these details, the exponent from
integrating the SH SF is $-1/4$. 

According to the GL model, at small $\varepsilon$ one expects
$\xi\sim\varepsilon^{-\nu}$ with $\nu=1/2$ \cite{Cross,CMT}. One also
expects the total power to vanish at onset in proportion to $\varepsilon$.
Thus, if Eq.~\ref{eq:SHSF} or Eq.~\ref{eq:squaredSHSF} gives the
shape of $S\left(k\right)$ correctly, then according to Eq.~\ref{eq:SH_expandedpower}
or Eq.~\ref{eq:squaredSH_expandedpower}, $P\sim\varepsilon^{\nu+\beta}$with
$\nu+\beta=1$ and $B\sim\varepsilon^{\beta}$ with $\beta=1/2$.

\section{\label{sub:SkMethods}Estimation of $S\left(k\right)$ from pattern images}
We found that the accuracy of $S\left(k\right)$, estimated from the Fourier power spectrum,
suffered greatly due to the finite size of the images of patterns that are available from experiment.  In applying the Fourier-transform (FT) method
to compute $S\left(k\right)$ from experimental data, 
we divided the images by a reference image taken below onset, applied a Kaiser-Bessel window \cite{KK66} to the divided images, and computed the magnitude squared of the FT. The azimuthal average of the squared magnitude of the FT yielded the SF $S\left(k\right)$.

We experimented
with three windowing functions: a square window, a Welch window, and
a Kaiser-Bessel window \cite{KK66}

\begin{equation}
w_{KB}\left(t\right)=\left\{ \begin{array}{cc}
\textrm{I}_{0}\left(\pi\alpha\sqrt{1-\left[t/\tau\right]^{2}}\right)/\textrm{I}_{0}\left(\pi\alpha\right) & \left|t\right|<\tau\\
0 & \left|t\right|>\tau\ .\end{array}\right.\label{eq:Kaiser-Bessel}\end{equation}

\noindent In Eq.~\ref{eq:Kaiser-Bessel} $t$ is the distance along one of the axes from the center, $\tau$ is the half-width of the image, and $\textrm{I}_{0}$ is the modified Bessel function of the first kind and order zero. The parameter $\alpha$ controls the rate at which the window drops off as the edge of the image is approached.
Since the data was two dimensional, the window was the product of
a one-dimensional window function in each direction.  Since the windowing function attenuates the signal around the
edges of the image, it reduces the total power.  To compensate for this we divided $S\left(k\right)$ by a constant
so that the total power in the final SF agreed with the total power of the raw image.

The specific windowing function did not greatly affect the result
for $\xi$. We ultimately settled on using a Kaiser-Bessel window
with $\alpha=2.5$ for the results given in this paper. Our major results,
namely the exponents $\nu$ and $\beta$ and the scaling collapse of the SF, were
independent of the windowing function.

We used a fast Fourier-transform (FFT) algorithm \cite{FFTW} capable of transforming images of arbitrary size  to compute
the SF so that no interpolation or zero padding was required to re-size
the image to an integer power of two. We found that interpolation
distorts the large-$k$ behavior of the SF because it smooths
out random noise that is present in the original signal. The use of
zero padding circumvents the smoothing of the noise and thus avoids
this distortion at large $k$, but instead inserts ringing at small
$k$. Due to the easy availability of FFT algorithms that can work
on any size image and the speed of modern computers, there is no reason
to sacrifice the behavior of the SF at either large or small $k$.
A minor tradeoff is that previous work utilized an aspect-ratio correction
to the image that removed a slight anisotropy in the frame grabber
\cite{Hu}. Because this correction requires interpolation of the
raw image we avoided using it. We expected only a very slight error
by doing so because the raw images are nearly square.  However, in the $\Gamma=36$ sample
there was a radial distortion, due to optical aberration, that was strong enough to warrant its removal in order to avoid
a systematic error in the length scale.  We did not investigate the large $k$ behavior in that sample, so
we expect no significant problem from the distortion correction.

Even after dividing the experimental images by an optical background, the resulting SF still contained the non-deterministic part of the background
spectrum present in the images taken below onset. Since fluctuations \cite{WAC95,OA03} are too feeble to be detected for the parameters of the present
experiment, we attributed this background signal primarily to electronic
noise and subtracted a background, determined below onset, from the SF above onset. As a result, for the case of
experimental data,
the analysis that follows is applied to the background subtracted
SF $\delta S\left(k\right)\equiv S\left(k\right)-S_{b}\left(k\right)$
where $S_{b}\left(k\right)$ is the SF averaged over many images below
the onset.  We also applied the FT to simulations of the SH domain-chaos model.  In that case there was
no need to divide by a reference image or to subtract electronic noise, so we used $S\left(k\right)$ computed directly
from the simulation images.

We attempted to overcome the finite image-size problem of the FT method by using the maximum entropy method (MEM) to estimate $S\left(k\right)$.
Although this method is commonly used for 1D data \cite{NR}, computing the spectrum of 2D data with this technique 
is still somewhat of an open problem \cite{MEM3}.  We implemented the algorithm detailed in Ref.~\cite{MEM1} 
which uses an iterative
method to arrive at the power-spectrum estimate. Figure 4 in Ref.~\cite{MEM1} provides a detailed flowchart 
of the MEM algorithm, which we followed precisely. The MEM provides the power spectrum as an expansion of the form

\begin{equation}
S\left(k_x,k_y\right)=\frac{1}{F\left[\lambda\left(n_1,n_2\right)\right]},
\label{eq:MEM_expansion}\end{equation}

\noindent where $\lambda \left(n_1,n_2 \right)$ are the coefficients of the expansion and $F\left[\cdots\right]$ is the discrete
Fourier transform.  
Because the MEM provides an expansion containing sines and cosines
in the denominator, in contrast with the FT where sines and cosines are in the numerator, the resulting
power spectrum may more accurately represent sharp peaks because it may contain poles \cite{NR}.  It
was also shown previously \cite{MEM2} that the MEM is much less sensitive to short data 
lengths than standard Fourier analysis.

The algorithm 
in Ref.~\cite{MEM1} depends on an accurate estimate of the auto-correlation of the data.  A central section
is cut from the auto-correlation data and used in the iterative process.  For all the analysis in the present
work we used a region size of $45\times45$ data points, which corresponded to about $12d\times12d$ for the
$\Gamma=61.5$ sample and also for the SH simulation. The iterative process produces auto-correlation data that
is continued beyond this region according to the spectral-entropy maximization-criteria that defines the MEM. The
number of coefficients $\lambda\left(n_1,n_2\right)$ determines the size of the continued region.  In principle it
is possible to use a relatively small continued-region size, in order to reduce the required amount of computation,
and then embed the resulting $\lambda\left(n_1,n_2\right)$ into a larger region, setting higher order coefficients to
zero in order to produce a finely meshed power spectrum.  In practice we found that this approach, while advocated
in Ref.~\cite{MEM1}, did not always produce positive-definite power spectra and thus was not stable for our purposes.
Instead we ran the entire algorithm on a large $720\times720$ mesh for $\lambda\left(n_1,n_2\right)$.  Although
this was heavily computation intensive, it did provide a reliable power spectrum estimate after a sufficient number
of iterations.  On our modest fleet of 2 GHz PowerMac G5s, we could run roughly 200 iterations per minute per CPU.
Typically we needed 50,000 to 1 million iterations for convergence, depending on the $\varepsilon$ step, 
making the processing of all the data in the present work a fairly massive undertaking.

The number of iterations depended on 
the auto-correlation error $\epsilon_0$, as defined by Eq.~25 in Ref.~\cite{MEM1}, which was reduced to a 
specified level.  In the case of the SH simulation data discussed in Sect.~\ref{sub:SimSH}, we found that
$\epsilon_0=10^{-5}$ was sufficiently small to give convergence of the SF.  Table \ref{tab:converge_test}
shows the results for $\nu$ and $\beta$ in the convergence test.  By $\epsilon_0=10^{-5}$, $\nu$ and $\beta$ have
nearly reached a constant.       	
In the case of the experimental data discussed in Sect.~\ref{sub:Experimental_Measurement_xi},
we found that we only needed $\epsilon_0=10^{-3}$ for satisfactory convergence.  This is likely due
to the better statistics for the auto-correlation function in the case of the experiment due to the fact
that we averaged the auto-correlation function over 4096 images per $\varepsilon$ step in the
experiment, but only 256 images per $\varepsilon$ step in the simulation.

\begin{table}
\caption{\label{tab:converge_test}Convergence test for SH simulation data with $\Gamma^*=150$.}
\begin{ruledtabular}
\begin{tabular}{ccc}
  $\epsilon_0$ & $\nu$ & $\beta$\\
  $10^{-3}$ & 0.858 & 0.277\\	  
  $10^{-4}$ & 0.619 & 0.455\\	  
  $10^{-5}$ & 0.555 & 0.496\\	  
  $10^{-6}$ & 0.533 & 0.506\\	  
\end{tabular}
\end{ruledtabular}
\end{table}

We applied the MEM to the pattern images in a similar way that the FT was used. We averaged the
auto-correlation over many images in order to reduce statistical fluctuations.  Then the MEM was applied to this
averaged auto-correlation to yield a single estimate of $S\left(k_x,k_y\right)$ at each temperature step.  
This
is in contrast to the Fourier analysis where we computed an $S\left(k_x,k_y\right)$ for each image, directly from
the Fourier power spectrum, and then averaged
them together to get an averaged spectrum.  Since the Fourier transform is linear, this technique is algebraically
equivalent to making a single Fourier transform of an averaged auto-correlation function, similar to what was done
for the MEM case.  Due to the iterative nature of the MEM algorithm we used, it was prohibitively slow to compute an
$S\left(k_x,k_y\right)$ for each image individually.

We used the SF to determine the correlation length. However, we modified the analysis scheme compared
to earlier work \cite{HEA1,MBCA93}, in light of the results presented
in Sect.~\ref{sub:Analytic_Moments}, to avoid using numerical moments.
We determined $\xi$ by fitting each of the functions
in Eqs.~\ref{eq:SHSF}-\ref{eq:LorentzianSF}, multiplied by the shadowgraph transfer function from Refs.~\cite{BBMTHCA96,TC03}, to the data near the
peak of the SF averaged over all images at a particular $\varepsilon$.

\section{\label{sub:SimSH}Finite image size effect in simulation of SH model for domain chaos}

\begin{figure}
\begin{center}\includegraphics[width=3.33in, keepaspectratio]{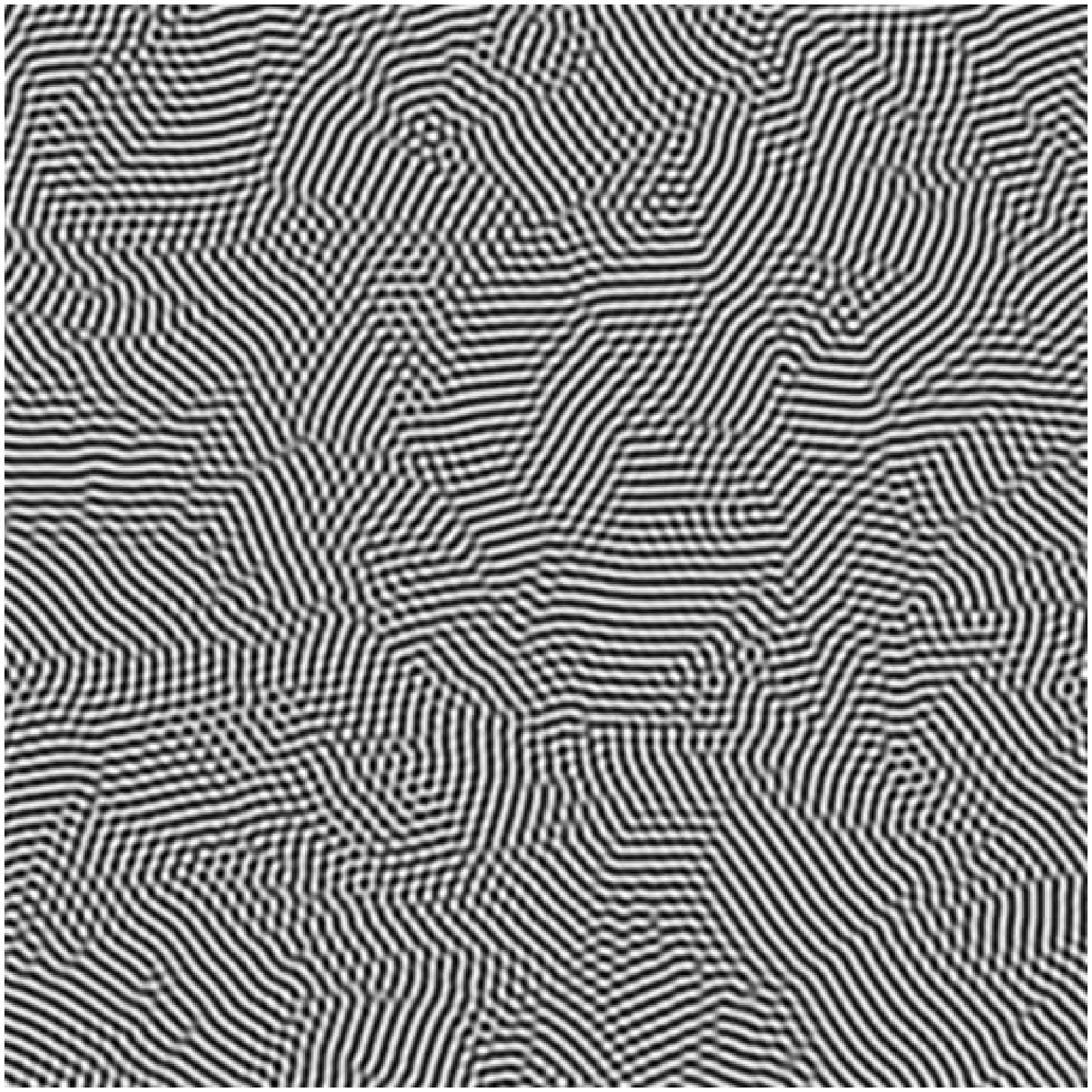}\end{center}
\caption{\label{cap:psiexample}A solution of Eq.~\ref{eq:SH_domain_chaos}  for $\tilde{\varepsilon}=0.12$.  This is
a $512\times512$ ($\Gamma^*=150$) cutout from the center of a $1024\times1024$ image with $\Gamma^*=300$. A movie
is available \cite{MOV2}.}
\end{figure}

The finite size of the pattern images has a significant impact on the accuracy
of a measurement of $\xi$ from $S\left(k\right)$.  To study this effect, 
we ran simulations using the Swift-Hohenberg model for domain chaos \cite{CMT}.  We utilized the
algorithm in Ref.~\cite{CMT} and periodic boundary conditions to solve the equation 

\begin{widetext}
\begin{equation}
\partial_{t}\psi=\tilde{\varepsilon}\psi-\left(\nabla^{2}+1\right)^{2}\psi-g_{1}\psi^{3}+g_{2}\hat{z}\cdot\nabla\times\left[\left(\nabla\psi\right)^{2}\nabla\psi\right]+g_{3}\nabla\cdot\left[\left(\nabla\psi\right)^{2}\nabla\psi\right]
\label{eq:SH_domain_chaos}
\end{equation}
\end{widetext}

\noindent for $\psi$, a field that can be used to model the temperature of the convection sample at the midplane. 
Figure \ref{cap:psiexample} shows an example of $\psi$.  The time step for numerical integration was $0.1$.
The initial condition for $\psi$ was a random grid of straight-roll
patches.  At each $\tilde{\varepsilon}$, $10000$ warmup time steps were performed followed by $256$ snapshots of $\psi$ recorded at an interval of $1250$ time steps.  The pixel spacing was chosen to reflect a non-unity sample thickness, unlike the    
choice in Ref.~\cite{CMT}, in order for the wave number and $\xi$ to be         
nearer the values in the experiment.  The choice of this constant has no        
effect on the value of $\nu$, $\beta$, or $\mu$, thus it does not affect our    
main conclusions.

\begin{figure}
\begin{center}\includegraphics[width=3.33in, keepaspectratio]{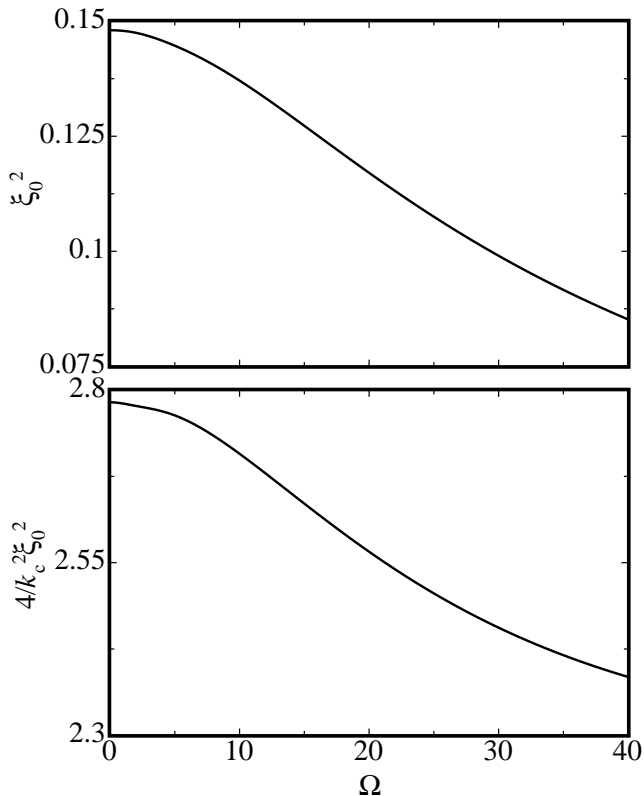}\end{center}
\caption{\label{cap:xi02and}Top figure: curvature $\xi_0^2$ of the neutral curve.  Bottom figure: $4/k_c^2\xi_0^2$.}
\end{figure}

The control parameter
in the simulation is related to the experimental control parameter by $\tilde{\varepsilon}=\left(4/k_c^2\xi_0^2\right)\varepsilon=2.60\varepsilon$ \cite{CH93},
where $k_c$ is the critical wave number and $\xi_0$ is  the curvature of the neutral curve
\cite{SCEQ}.  The numerical value $2.60$ corresponds
to $\Omega=17.5$.  Figure \ref{cap:xi02and} shows both $\xi_0^2$ and $4/k_c^2\xi_0^2$ as a function of $\Omega$ as computed numerically from the neutral curve. We note that the exact
value of $4/k_c^2\xi_0^2$ has no effect on the results for $\xi$, $B$, $\nu$, $\beta$, $f$, or $\mu$, and only serves to adjust the range
of $\varepsilon$. 

The parameters $g_1$, $g_2$, and $g_3$ control the stability balloon and can be chosen
to model a specific $\Omega$ and K\"uppers-Lortz angle $\theta_{KL}$.
In the present work we used $g_1=1$, $g_2=-2.4534$, and $g_3=0.522$ which
are the same parameters used in Fig.~1b of Ref.~\cite{CMT} and which correspond to $\theta_{KL}=51^\circ$ and
roughly $\Omega\simeq17.5$ for the Prandtl number in the present experiment. Note that Eq.~\ref{eq:SH_domain_chaos} 
assumes infinite Prandtl number, so this rough value of $\Omega$ comes from comparing
$\Omega/\Omega_c$, where $\Omega_c$ is Prandtl number dependent.

As a check on the accuracy of our implementation of the solver algorithm we attempted to reproduce Fig.~7a
of Ref.~\cite{CMT}.  Our simulation agreed within 2\% for the data points at $\tilde{\varepsilon}=0.1$, $0.2$, and $0.3$ 
in that figure, but differed by about 27\% for the data point at $\tilde{\varepsilon}=0.03$ for an unknown reason.
Nevertheless, we are confident in the correctness of our simulation and in the results to follow.

In order to determine the effect of the finite image size on the measured value of $\xi$ we ran
simulations of image size $1024\times1024$ corresponding to an image aspect ratio $\Gamma^*=300$,
where $\Gamma^*$ refers to the horizontal width of the image.  This is in contrast to the aspect
ratio $\Gamma$ which refers to the physical aspect ratio of the convection sample.  From these large
simulations we cut various sized center sections and computed the SF using both the FT and the MEM following
the procedure described in Sect.~\ref{sub:SkMethods}.
We fit the SH function and the squared SH function to the SF.  

\begin{figure}
\begin{center}\includegraphics[width=3.33in, keepaspectratio]{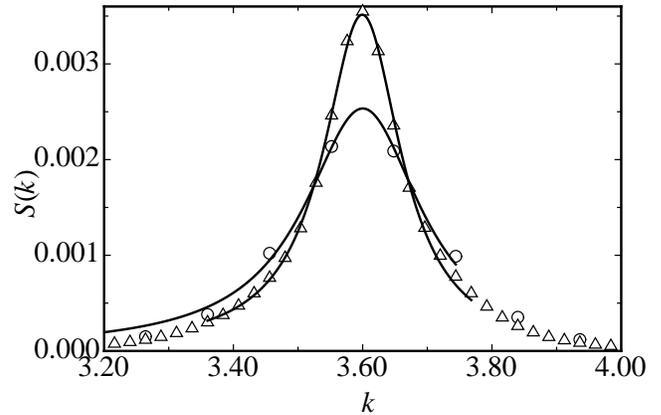}\end{center}
\caption{\label{cap:simSFfft_120}Structure factor (SF) from simulations using Eq.~\ref{eq:SH_domain_chaos} with $\tilde{\varepsilon}=0.12$,
computed with the Fourier-transform method.  Both data sets
are from the same simulation except with a different size center-cutout used to compute the SF.	
Triangles: $\Gamma^*=300$.  Circles: $\Gamma^*=75$. 
Solid lines: Fits of Eq.~\ref{eq:SHSF} to the data over the range $k\pm3/\xi$.}
\end{figure}

\begin{figure}
\begin{center}\includegraphics[width=3.33in, keepaspectratio]{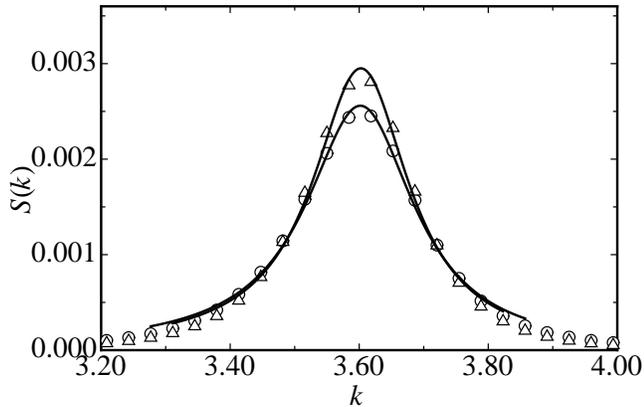}\end{center}
\caption{\label{cap:simSFmem_120}Structure factor (SF) from simulations using Eq.~\ref{eq:SH_domain_chaos} with $\tilde{\varepsilon}=0.12$,
computed with the maximum entropy method (MEM).  Both data sets
are from the same simulation except with a different size center-cutout used to compute the SF.	
Triangles: $\Gamma^*=300$.  Circles: $\Gamma^*=75$. 
Solid lines: Fits of Eq.~\ref{eq:SHSF} to the data over the range $k\pm3/\xi$.}
\end{figure}

The choice of $\Gamma^*$ strongly affected the shape of the SF when computed with the FT.  
Figure \ref{cap:simSFfft_120} compares the SF of the same data for two different values of $\Gamma^*$.
The same data analyzed with the MEM, as shown in 
Fig.~\ref{cap:simSFmem_120}, is affected slightly by $\Gamma^*$, but it is not nearly as sensitive as the FT.
It is important to note that the sharpest SF peak is not necessarily the best. 
The $\Gamma^*=300$ SF from the FT is sharper than either of the MEM peaks, but it is not at all consistent with
its shorter data length relative, $\Gamma^*=75$.  In contrast, the MEM SFs land nearly on top of each other when
comparing $\Gamma^*=300$ and $\Gamma^*=75$.  The ability to accurately represent the SF peak for relatively small image sizes
is crucial because system sizes as large as $\Gamma^*=300$ are experimentally inaccessible.

\begin{figure}
\begin{center}\includegraphics[width=3.33in, keepaspectratio]{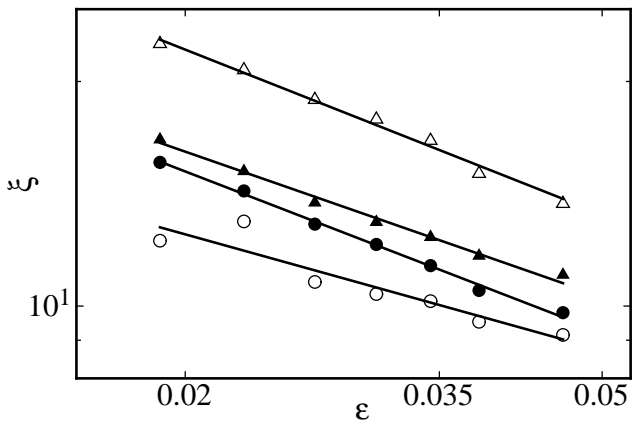}\end{center}
\caption{\label{cap:xis_sim_SH}Measurements of $\xi$ from fits of Eq.~\ref{eq:SHSF} to many SFs like
those shown in Figs.~\ref{cap:simSFfft_120} and \ref{cap:simSFmem_120}. Open symbols: from the FT method.
Solid symbols: from the MEM. Triangles: $\Gamma^*=300$.  Circles: $\Gamma^*=75$.
Solid lines: Fits of the power law $\xi\propto\varepsilon^{-\nu}$ to the data over the range shown.}
\end{figure}

From fitting the SF for both the FT and MEM, we computed $\xi$ for many $\varepsilon$ and $\Gamma^*$ as shown 
in Fig.~\ref{cap:xis_sim_SH}.  Once again
we found that the MEM provides significantly more consistent results for vastly different $\Gamma^*$.  In the case
of the FT results, it is critical to note that increasing $\Gamma^*$ does not simply increase $\xi$ due to the
larger data length.  It also increases the slope of $\xi$ {\it vs.} $\varepsilon$ on logarithmic scales, thus affecting the measured $\nu$.  Since we wish to
measure $\nu$ accurately enough to compare with the model prediction $\nu=1/2$, this is a discouraging outcome. 

\begin{figure}
\begin{center}\includegraphics[width=3.33in, keepaspectratio]{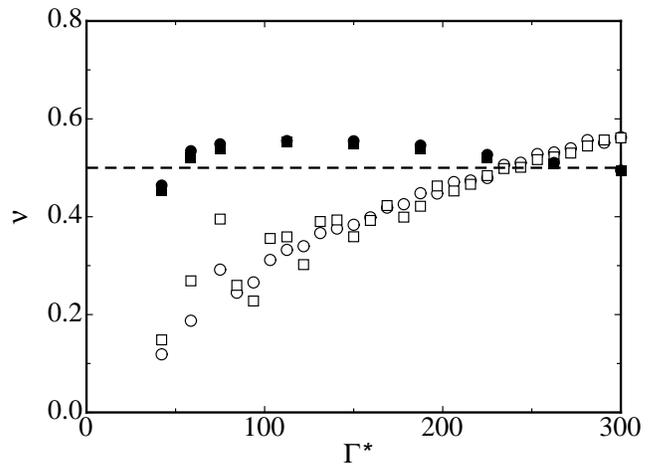}\end{center}
\caption{\label{cap:nusim}Results for $\nu$ from fits like those shown in Fig.~\ref{cap:xis_sim_SH}. 
Solid symbols: from the MEM.  Open symbols: from the FT method.  Circles: $\xi$ was determined 
by fitting the SH SF Eq.~\ref{eq:SHSF} to the data.  Squares: $\xi$ was determined by fitting a squared 
SH SF Eq.~\ref{eq:squaredSHSF} to the data.  Dashed line: the $\nu=1/2$ prediction from model equations.}
\end{figure}

We ascertained the effect of image size on the accuracy of $\nu$ by computing $\nu$ as a function of $\Gamma^*$
as shown in Fig.~\ref{cap:nusim}.  The two lowest $\Gamma^*$ points in that figure correspond to the available image size in the experimental samples discussed in the present work.  The $\nu$ values from the MEM and FT method both
approached nearly the same asymptotic value at large $\Gamma^*$.  Fortunately, the result from the MEM has nearly reached this
asymptote at even the smallest $\Gamma^*$ indicating that it may be reliably used for measuring $\nu$ even in the
case of small images.  Not only is the FT measurement of $\nu$ much too low at the smallest $\Gamma^*$, but it
also contains rather large fluctuations at those values.  At the largest $\Gamma^*$ it is almost as good as the MEM, but
that does not help for the analysis of experiments which are limited roughly to $\Gamma^* < 100$ \cite{BSCA05}.

\begin{figure}
\begin{center}\includegraphics[width=3.33in, keepaspectratio]{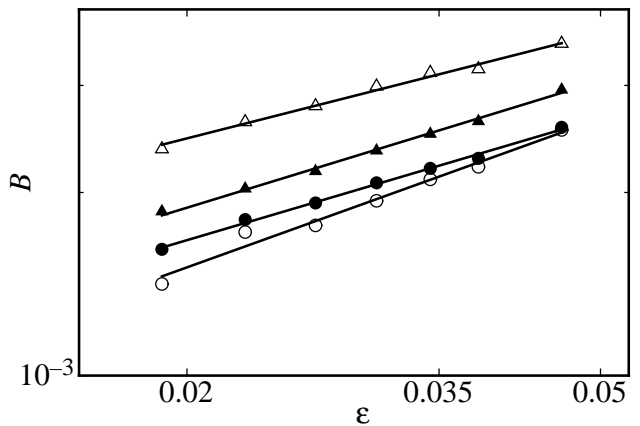}\end{center}
\caption{\label{cap:Bexamples_SH}Results for $B$ from fits of Eq.~\ref{eq:SHSF} to many SFs like
those shown in Figs.~\ref{cap:simSFfft_120} and \ref{cap:simSFmem_120}.
Open symbols: from the FT method. Solid symbols: from the MEM.
Triangles: $\Gamma^*=300$.  Circles: $\Gamma^*=75$.
Solid lines: Fits of the power law $B\propto\varepsilon^\beta$ to the data over the same range as shown in Fig.~\ref{cap:xis_sim_SH}.}
\end{figure}

We observed a similar finite image-size effect for the scaling of $B$ with $\varepsilon$.  As discussed in Sect.~\ref{sub:Analytic_Moments}
we expect that $B\sim\varepsilon^{1/2}$ provided that $\xi$ obeys the scaling in the
amplitude model, i.e. $\xi\sim\varepsilon^{-1/2}$.  Our SF fits that yielded $\xi$ also gave $B$.  
Figure \ref{cap:Bexamples_SH} shows some examples of $B$ for the same data as shown in Fig~\ref{cap:xis_sim_SH}. 

\begin{figure}
\begin{center}\includegraphics[width=3.33in, keepaspectratio]{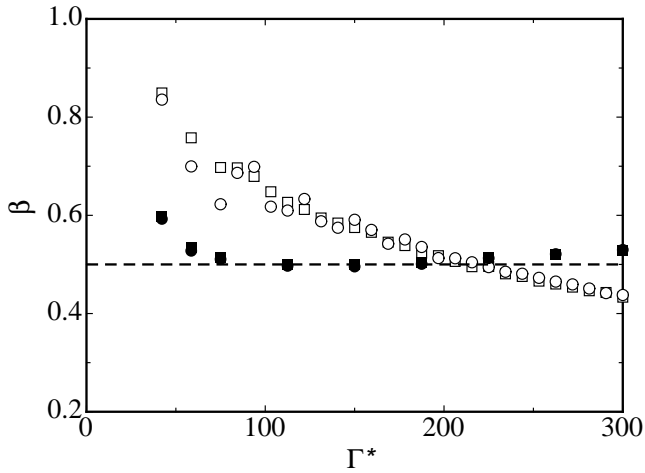}\end{center}
\caption{\label{cap:betasim}Results for $\beta$ from fits like those shown in Fig.~\ref{cap:Bexamples_SH}. 
Solid symbols: from the MEM.  Open symbols: from the FT method.  Circles: $\xi$ was determined 
by fitting the SH SF Eq.~\ref{eq:SHSF} to the data.  Squares: $\xi$ was determined by fitting the squared 
SH SF Eq.~\ref{eq:squaredSHSF} to the data.  Dashed line: the $\beta=1/2$ prediction from model equations.}
\end{figure}

We extracted a scaling exponent $\beta$ by fitting the equation $B\propto\varepsilon^\beta$ to the data. As indicated
by Fig.~\ref{cap:betasim}, the result is similar to the finite image-size effect on $\nu$.  At the smallest 
$\Gamma^*$ values, the MEM results exhibit a slight $\Gamma^*$ dependence, but they quickly reach a nearly constant value as $\Gamma^*$ increases.  The FT method
suffers from a strong $\Gamma^*$ dependence.  However, it is quite satisfying that for $\Gamma^*$ near the
experimental values, we obtained $\nu\simeq0.25$ and $\beta\simeq0.75$ from the FT method, in agreement with what we (and others
previously) observed in the experiment.

\begin{figure}
\begin{center}\includegraphics[width=3.33in, keepaspectratio]{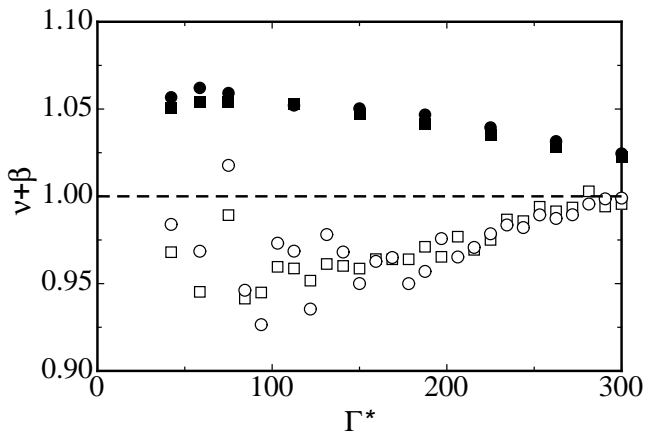}\end{center}
\caption{\label{cap:nupbetasim}Values of $\nu+\beta$ from combining the results shown in Figs.~\ref{cap:nusim} and \ref{cap:betasim}. 
Solid symbols: from the MEM.  Open symbols: from the FT method.  Circles: $\xi$ and $B$ determined 
by fitting the SH SF Eq.~\ref{eq:SHSF} to the SFs.  Squares: $\xi$ and $B$ determined by fitting the squared 
SH SF Eq.~\ref{eq:squaredSHSF} to the SFs.  Dashed line: $\nu+\beta=1$.}
\end{figure}

Figure \ref{cap:nupbetasim} shows the results for $\nu+\beta$ for both the MEM and the FT method.  Both are close to the expected result $\nu+\beta=1$, and approach it even more closely as $\Gamma^*$ increases.

\begin{figure}
\begin{center}\includegraphics[width=3.33in, keepaspectratio]{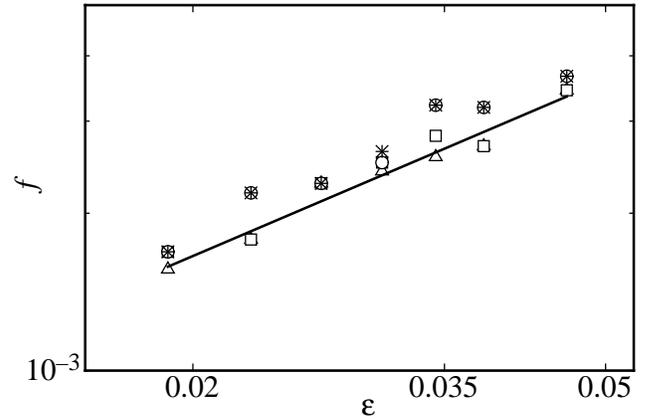}\end{center}
\caption{\label{cap:timescale}Determinations of the characteristic frequency scale $f$ of the domain chaos from the auto-correlations of time-angle
plots of the SFs computed with the FT method.
Triangles: $\Gamma^*=300$.  Squares: $\Gamma^*=150$. Circles: $\Gamma^*=75$.  Crosses: $\Gamma^*=58.6$.
Pluses: $\Gamma^*=42.2$.
Solid line: Fit of the power law $f\propto\varepsilon^{\mu}$ to the data over the same range as shown in Fig.~\ref{cap:xis_sim_SH}.}
\end{figure}

Unlike for $\xi$ and $B$, the finite image size does not interfere with the determination of the characteristic frequency scale $f$ of the domain chaos (the domain precession frequency) 
when the FT method is used to compute the SF.  We computed $f$ using the method explained in Ref.~\cite{HPAE98}.  
We extracted a time scale from the auto-correlation of the angle-time plot of the radially averaged SF.  This yielded
the results for $f$ shown in Fig.~\ref{cap:timescale}.  All of the $f$ data in that figure land
nearly on top of each other regardless of $\Gamma^*$.  Since the measurement of $f$ depends more on the position
of the peak in the two-dimensional SF than on its width, the FT method is able provide enough resolution to accurately measure
$f$ at smaller $\Gamma^*$.

\begin{figure}
\begin{center}\includegraphics[width=3.33in, keepaspectratio]{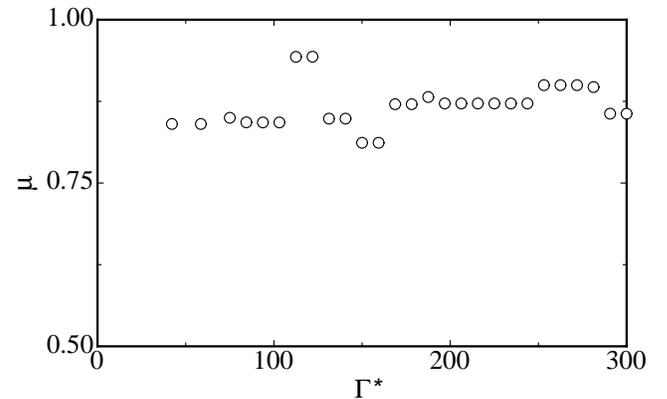}\end{center}
\caption{\label{cap:musim}Measurements of $\mu$ from fits, like those shown in Fig.~\ref{cap:timescale},
to data for $f$ computed from SFs with the FT method.} 
\end{figure}

We fit a power law $f\propto\varepsilon^\mu$ to $f$ because the GL model predicts such a power law with $\mu=1$.
Figure \ref{cap:musim} shows that this exponent is independent of $\Gamma^*$.  However, the result $\mu \simeq 0.9$ is somewhat lower
that the predicted value. We do not know the reason for this. However, the value we found for $\mu$ is much larger than the experimental value $\mu \simeq 0.6$ \cite{HEA1}. Recently it was found \cite{BSCA05} that this difference is caused by the centrifugal force which influences the experiment but is omitted in Eq.~\ref{eq:SH_domain_chaos}. 

\section{\label{sub:Experimental_Measurement_xi}New experimental measurements}

We acquired new data using an apparatus described previously \cite{BBMTHCA96}.
There were two samples which both used  compressed SF$_6$ gas.  The primary sample
was at a pressure of 20.00 bars
and mean temperature of $38.00^{\circ}\textrm{C}$ where the Prandtl
number $\eta/\left(\rho\kappa\right)$ ($\kappa$ is the thermal diffusivity)
was $0.87$. The aspect ratio $\Gamma\equiv D/\left(2d\right)$ ($D$
is the sample diameter) was 61.5. A secondary sample with $\Gamma=36$, which is discussed briefly in this work, was
pressurized at 12.34 bars with a mean temperature of $38.00^{\circ}\textrm{C}$ and the Prandtl number was $0.82$.

\begin{figure}
\begin{center}\includegraphics[width=3.33in, keepaspectratio]{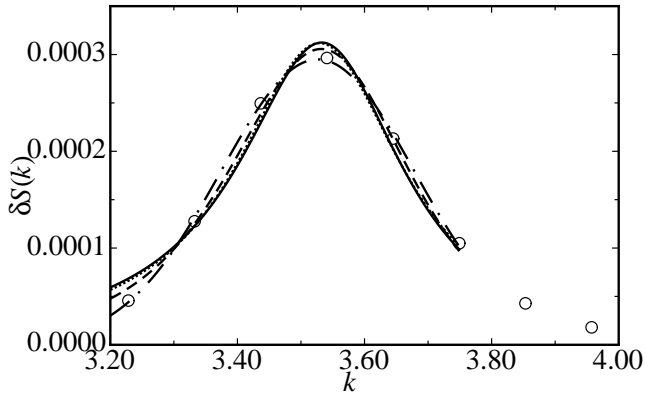}\end{center}
\caption{\label{cap:SF_FT_fitting}Fits of Eqns.~\ref{eq:SHSF}-\ref{eq:LorentzianSF}
to the experimental SF computed with the FT for $\Omega=17.7$ and $\varepsilon = 0.05$. Solid line: fit of SH SF Eq.~\ref{eq:SHSF}. Dashed line: fit of squared SH SF Eq.~\ref{eq:squaredSHSF}. Dashed-dotted line: fit of Gaussian SF Eq.~\ref{eq:GaussianSF}. Dotted line: fit of Lorentzian SF Eq.~\ref{eq:LorentzianSF}.  All fitting functions are multiplied by the shadowgraph transfer function.}
\end{figure}

\begin{figure}
\begin{center}\includegraphics[width=3.33in, keepaspectratio]{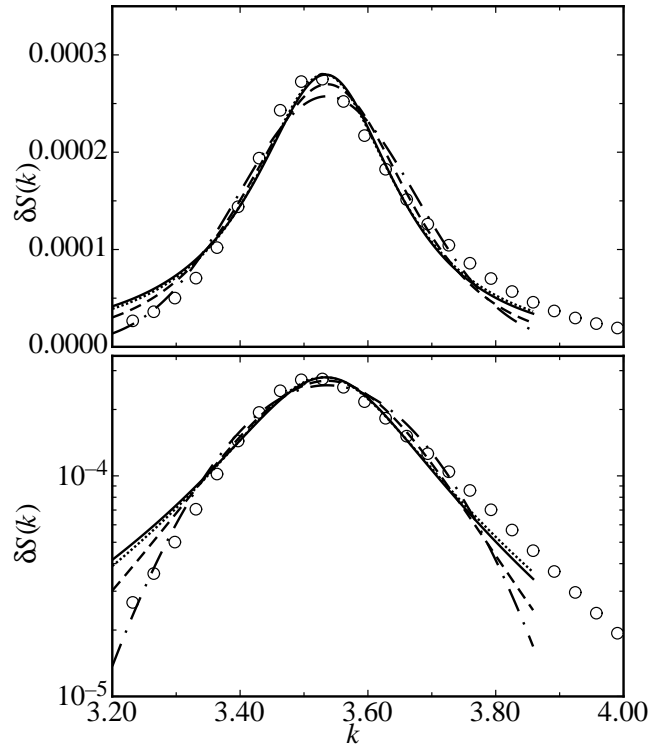}\end{center}
\caption{\label{cap:SF_MEM_fitting}Fits of Eqns.~\ref{eq:SHSF}-\ref{eq:LorentzianSF}
to the experimental SF computed with the MEM for $\Omega=17.7$ and $\varepsilon = 0.05$. Top figure: linear-linear plot. Bottom figure: log-linear plot. Solid line: fit of SH SF Eq.~\ref{eq:SHSF}. Dashed line: fit of squared SH SF Eq.~\ref{eq:squaredSHSF}. Dashed-dotted line: fit of Gaussian SF Eq.~\ref{eq:GaussianSF}. Dotted line: fit of Lorentzian SF Eq.~\ref{eq:LorentzianSF}.  All fitting functions are multiplied by the shadowgraph transfer function.}
\end{figure}

We determined $\xi$ by fitting each of the functions
in Eqs.~\ref{eq:SHSF}-\ref{eq:LorentzianSF}, multiplied by the shadowgraph transfer function from Refs.~\cite{BBMTHCA96,TC03}, to the data near the
peak of the SF. Typically we used a range $k_0 \pm 3/\xi$ where the initial guess of $\xi$ and $k_0$ for the fitting range was estimated by Eq.~\ref{eq:SH_expandedpower} using the zeroth numerical moment, peak value, and peak position.  The fit was not very sensitive to the range of $k$ provided that the range included a sufficient number of points. Figure \ref{cap:SF_FT_fitting} shows that for the FT method, all of the functions provide a good fit near the half height, where $\xi$ is determined, and near the peak where the fits give the value of $B$. Figure \ref{cap:SF_MEM_fitting} shows similarly good fitting 
results for the SF estimated with the MEM.

\begin{figure}
\begin{center}\includegraphics[width=3.33in, keepaspectratio]{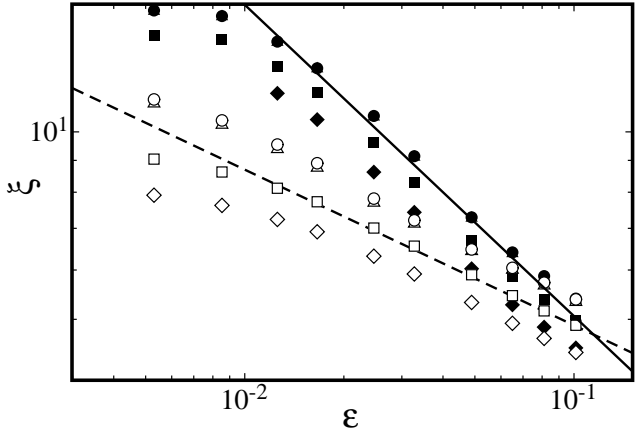}\end{center}
\caption{\label{cap:xi_fitting}Measurements of $\xi$ by fitting Eqns.~\ref{eq:SHSF}-\ref{eq:LorentzianSF}
to the experimental SF for $\Omega=17.7$. Open symbols: SF was computed using a Kaiser-Bessel
window with $\alpha=2.5$. Solid symbols: SF was computed with the MEM.
Dashed line: $\varepsilon^{-1/4}$.
Solid line: $\varepsilon^{-1/2}$. Circles: $\xi$ from fitting of
SH SF. Squares: $\xi$ from fitting of squared SH SF. Diamonds: $\xi$
from fitting of Gaussian SF. Triangles: $\xi$ from fitting of Lorentzian
SF.}
\end{figure}

Figure \ref{cap:xi_fitting}
shows the effect of the fitting function on the dependence of $\xi$ on $\varepsilon$ for $\Omega = 17.7$.
Although each of the fitting functions gave slightly different values
for $\xi$, all of them gave nearly the same value for $\nu$, with $\nu\simeq0.25$ from the FT method and $\nu\simeq0.5$
from the MEM. This suggests that the function fitting
is a robust method for measuring $\xi$ because knowledge of the exact
functional form of the SF is not required to yield a consistent measurement
for $\xi$. In spite of this consistency for $\nu$, we concluded that some fitting
functions are better than others as will be seen below.

\begin{figure}
\begin{center}\includegraphics[width=3.33in, keepaspectratio]{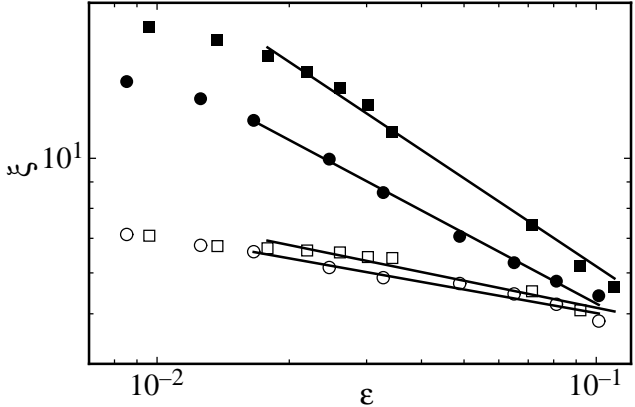}\end{center}
\caption{\label{cap:xigamcompare}Comparison of $\xi$ for
$\Omega=17.7$ and both $\Gamma=36$ and $\Gamma=61.5$, measured with SFs computed from $44d\times44d$ square cutouts
using both the MEM and the FT method. Squares: $\Gamma=36$.
Circles: $\Gamma=61.5$. Open symbols: FT. Solid symbols: MEM. For $\Gamma=36$ the
fitting range was $0.018\leq\varepsilon\leq0.11$, MEM: $\nu=0.57$,
FT: $\nu=0.17$. For $\Gamma=61.5$ the fitting range was $0.017\leq\varepsilon\leq0.10$,
MEM: $\nu=0.45$, FT: $\nu=0.15$.}
\end{figure}

The results from the MEM are in much better agreement, than the results from the FT method, with the prediction $\nu=1/2$ from the GL model. This difference is even more dramatic when considering a smaller-sized cutout region
from the data.  Figure \ref{cap:xigamcompare} shows $\xi$ for two different $\Gamma$, where the analysis has been applied
to a central region of identical size in either case.  Although in the case of $\Gamma=61.5$ the data shown in Fig.~\ref{cap:xigamcompare}
is from the same raw data as shown in Fig.~\ref{cap:xi_fitting}, the FT method gives a somewhat smaller value of $\nu$
because of the reduced image size of the smaller cutout region.  The MEM also suffers some minor decrease in the
value of $\nu$, as this cutout size corresponds to the smallest $\Gamma^*$ data point in Fig.~\ref{cap:nusim}.
Although only barely noticeable in the FT result, the MEM clearly shows that $\xi$ is larger in the case of
$\Gamma=36$ as compared to $\Gamma=61.5$.  A likely explanation is the increased influence of the centrifugal force in
the larger sample. The presence of the centrifugal force has been shown to decrease the domain size \cite{BSCA05}.

\begin{figure}
\begin{center}\includegraphics[width=3.33in, keepaspectratio]{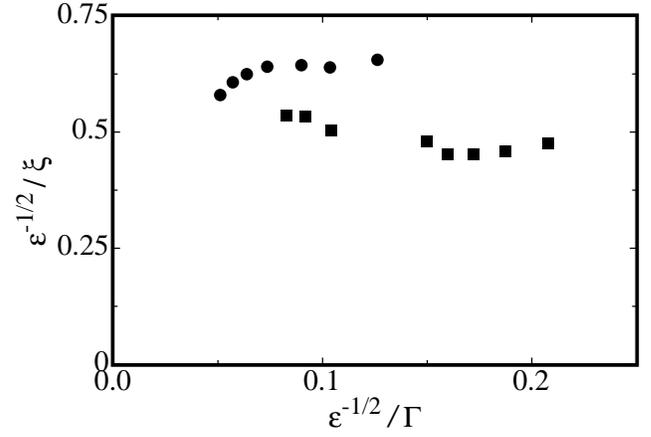}\end{center}
\caption{\label{cap:gamscale}$\varepsilon^{-1/2}$ scaled against $\xi$ and $\Gamma$ for  
$\Omega=17.7$ and both $\Gamma=36$ and $\Gamma=61.5$, measured with SFs computed from $44d\times44d$ square cutouts
using the MEM. Squares: $\Gamma=36$.
Circles: $\Gamma=61.5$. The data is the same as in Fig.~\ref{cap:xigamcompare} and is shown over the range of the
fits in that figure.}
\end{figure}

The diminishing slope at very small $\varepsilon$ shown in Figs.~\ref{cap:xi_fitting} and \ref{cap:xigamcompare}
may be indicative of a physical finite-size effect \cite{CLM01} due to the lateral boundary of the experimental samples.
However, over the range of fitting shown in Fig.~\ref{cap:xigamcompare}, no such effect is observed.  As a comparison
with the result in Fig.~3 of Ref.~\cite{CLM01}, Fig.~\ref{cap:gamscale} shows a plot of $\varepsilon^{-1/2}$ scaled with $\xi$ against
$\varepsilon^{-1/2}$ scaled with $\Gamma$.  The data is roughly constant over the range of $\varepsilon$ indicating agreement with
the GL prediction for $\nu$.  Unfortunately the data from two different $\Gamma$ do not collapse onto a single curve
as in Ref.~\cite{CLM01}.  This may be due to the effect of the centrifugal force, which was neglected in that work,
but is unavoidable in the present experiment.

\begin{figure}
\begin{center}\includegraphics[width=3.33in, keepaspectratio]{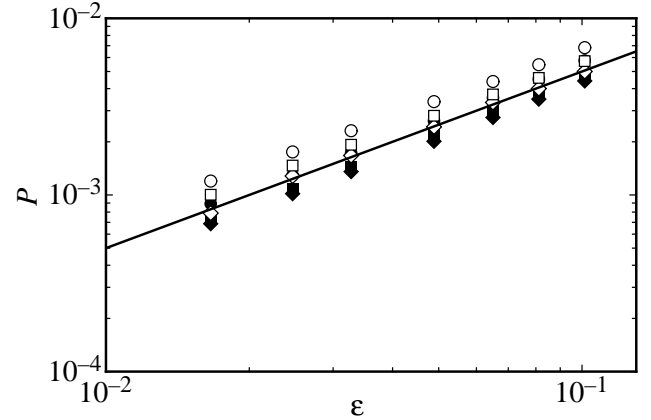}\end{center}
\caption{\label{cap:Total_Power_scaling}Total power for $\Omega=17.7$ from SFs computed with both the FT method and the MEM and given
by various methods. Solid line: $\varepsilon^{1}$. Open symbols: from the FT method.  Solid symbols: from the MEM.
Diamonds: power
from zeroth numerical moment of the SF data integrated over $k_0-3/\xi\le k\le k_0+3/\xi$. Circles: power given
by Eq.~\ref{eq:SH_exactpower} using the parameters acquired from
fitting the SH SF to the data. Squares: power given by Eq.~\ref{eq:squaredSH_exactpower}
using the parameters acquired from fitting the squared SH SF to the
data.}
\end{figure}

Although, in the case of the FT method, the scaling exponent $\nu$ did not agree with the GL model,
the scaling of the total power agreed for both the FT method and the MEM. Figure \ref{cap:Total_Power_scaling}
shows the total power for $\Omega = 17.7$ measured by several methods. The zeroth numerical
moment of the data did not perfectly agree with the total power computed
from Eq.~\ref{eq:SH_exactpower} or Eq.~\ref{eq:squaredSH_exactpower} but it was close.
In the case of those equations, the fit parameters $B$, $k_{0}$,
and $\xi$ (or $\xi_{s}$ for the squared SH) were used to calculate
$P$. Since Eqs.~\ref{eq:SH_exactpower} and \ref{eq:squaredSH_exactpower} represent the
total power over the range $0<k<\infty$ and the zeroth numerical moment was necessarily computed
over a finite range $k_0-3/\xi\le k\le k_0+3/\xi$, the slight discrepancy in total power is not surprising.  A numerical
integration of Eqs.~\ref{eq:SHSF} and \ref{eq:squaredSHSF} over a finite $k$ range yields a total power closer to the
numerical moment result, however, it is absent from Fig.~\ref{cap:Total_Power_scaling}
because it is too close to the other data points to be easily distinguishable.
In spite of this minor discrepancy, all three measurements of the total power are proportional to $\varepsilon$ as expected.

Since, in the case of the FT method, the dependence of $\xi$ on $\varepsilon$ differed from the
GL prediction of $\varepsilon^{-1/2}$, it is important to also investigate
the dependence of $B$ on $\varepsilon$. As shown in Fig.~\ref{cap:Total_Power_scaling},
the combination of $B$ and $\xi$ according to Eqs.~\ref{eq:SH_expandedpower}
and \ref{eq:squaredSH_expandedpower} yielded a total power that depended
on $\varepsilon$ in the predicted way for both the FT method and the MEM. 

\begin{figure}
\begin{center}\includegraphics[width=3.33in, keepaspectratio]{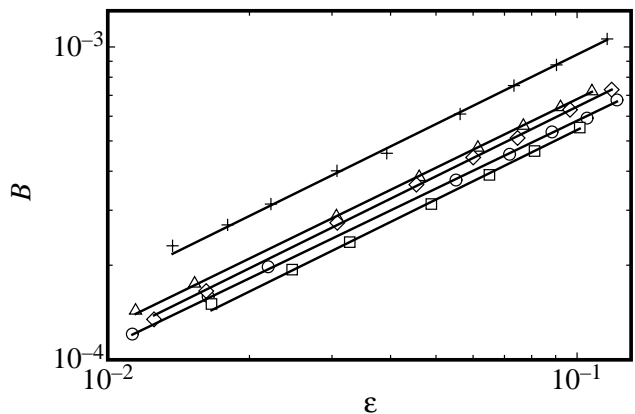}\end{center}
\caption{\label{cap:B_dataFT}$B$ determined by fitting the FT experimental SF to
the SH function. Solid lines: power law fits to the data to measure $\beta$. Pluses: $\Omega=15$.
Circles: $\Omega=16.25$. Squares: $\Omega=17.7$. Triangles: $\Omega=19.5$.
Diamonds: $\Omega=21.7$.}
\end{figure}

\begin{figure}
\begin{center}\includegraphics[width=3.33in, keepaspectratio]{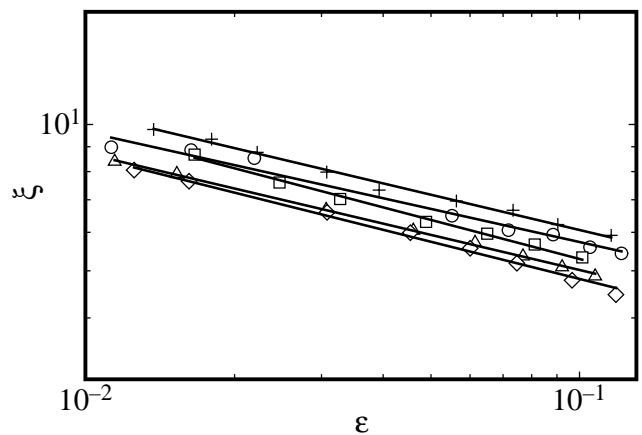}\end{center}
\caption{\label{cap:xi_dataFT}$\xi$ determined by fitting the FT experimental
SF to the SH function. Solid lines: power-law fits to the data to determine
$\nu$. Pluses: $\Omega=15$. Circles: $\Omega=16.25$. Squares: $\Omega=17.7$.
Triangles: $\Omega=19.5$. Diamonds: $\Omega=21.7$.}
\end{figure}

Figure \ref{cap:B_dataFT} shows
the result for $B$ measured by fitting the SH SF to the experimental
SF at all $\Omega$ values. The slight variation of the values of 
$\beta$ (slopes of the lines in the figure) did not depend  systematically on 
$\Omega$ and most likely it is due to experimental error. The power-law fits shown in the figure were over the same range as was used
to determine $\nu$ in Fig.~\ref{cap:xi_dataFT}. One sees that $B$ alone did not
depend on $\varepsilon^{1/2}$ as expected; instead $B\sim\varepsilon^{\beta}$
with, averaging over the results from all $\Omega$, $\beta\simeq0.73$
while the $\Omega$-averaged $\nu\simeq0.25$.
In other words, in the case of the FT method, $B$ and $\xi$ conspired to produce the
expected $P\sim\varepsilon^{\nu+\beta}$ with $\nu+\beta \simeq 1$.

\begin{figure}
\begin{center}\includegraphics[width=3.33in, keepaspectratio]{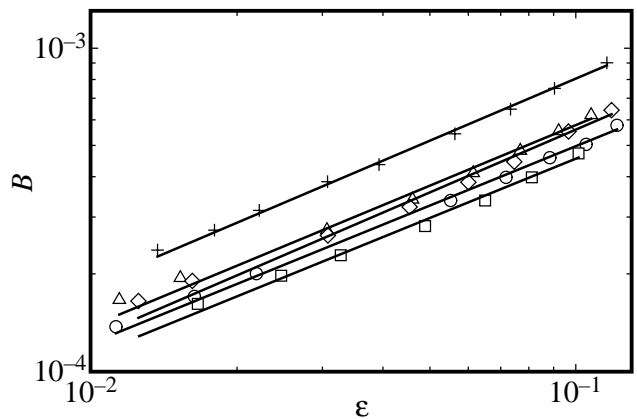}\end{center}
\caption{\label{cap:B_dataMEM}$B$ determined by fitting the MEM experimental SF to the SH function.
Solid lines: power-law fits to the data to determine $\beta$. Pluses: $\Omega=15$.
Circles: $\Omega=16.25$. Squares: $\Omega=17.7$. Triangles: $\Omega=19.5$.
Diamonds: $\Omega=21.7$.}
\end{figure}

\begin{figure}
\begin{center}\includegraphics[width=3.33in, keepaspectratio]{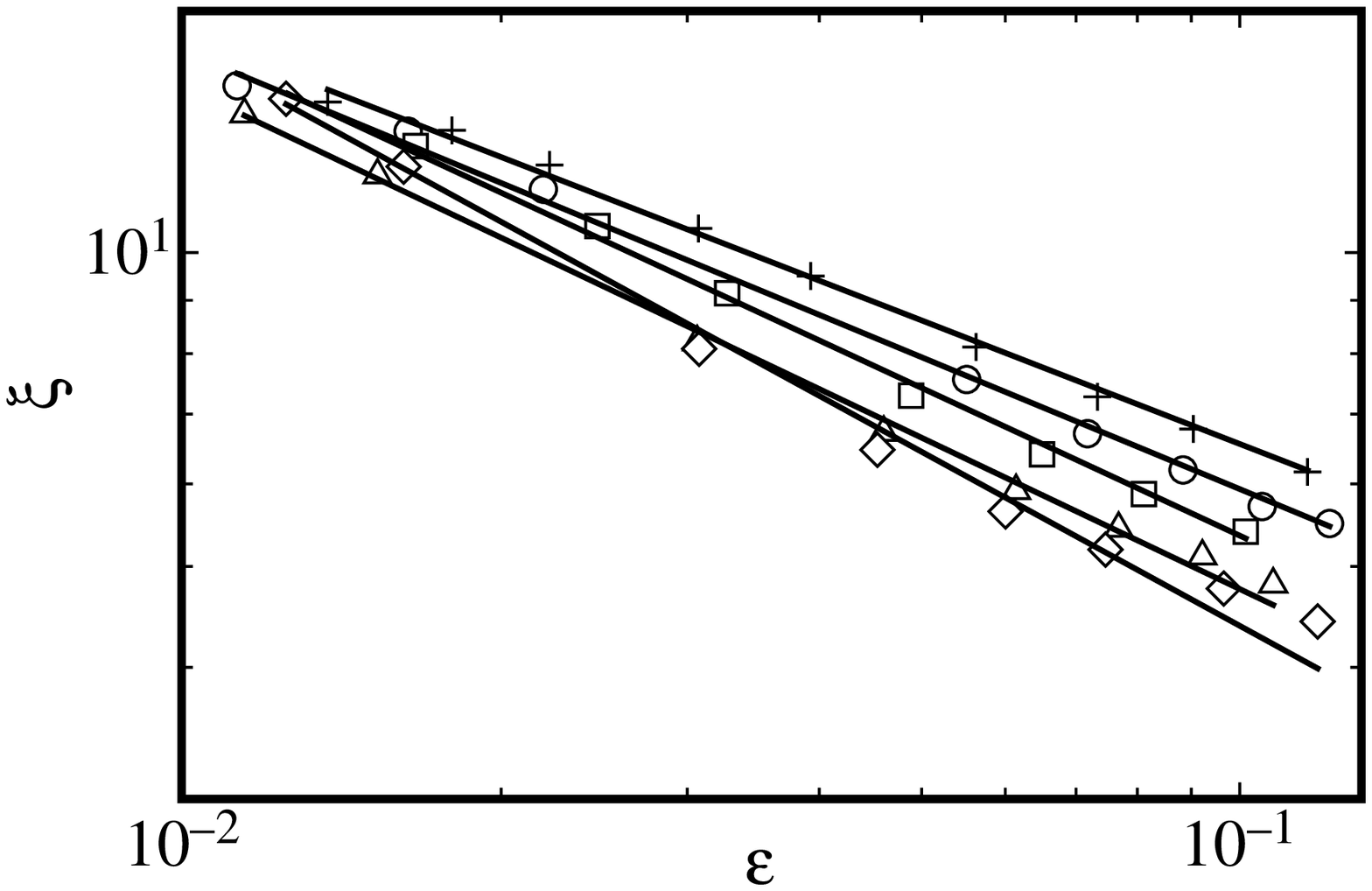}\end{center}
\caption{\label{cap:xi_dataMEM}$\xi$ determined by fitting the MEM experimental
SF to the SH function. Solid lines: power-law fits to the data to determine
$\nu$. Pluses: $\Omega=15$. Circles: $\Omega=16.25$. Squares: $\Omega=17.7$.
Triangles: $\Omega=19.5$. Diamonds: $\Omega=21.7$.}
\end{figure}

\begin{figure}
\begin{center}\includegraphics[width=3.33in, keepaspectratio]{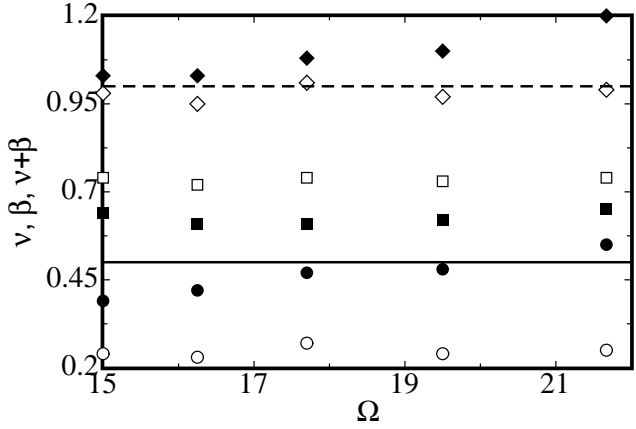}\end{center}
\caption{\label{cap:nu_Omega}Exponents from data shown in Figs.~\ref{cap:B_dataFT}-\ref{cap:xi_dataMEM}.
Open symbols: from the FT method.  Solid symbols: from the MEM.  Circles: $\nu$.  Squares: $\beta$.  Diamonds: $\nu+\beta$.
Solid line: $\nu=\beta=1/2$ prediction from GL model.  Dashed line: $\nu+\beta=1$.
}
\end{figure}

Figures \ref{cap:B_dataMEM} and \ref{cap:xi_dataMEM} show the analogous results from the MEM.  
The MEM yielded much more steeply sloped $\xi$ vs.~$\varepsilon$ curves as indicated by Fig.~\ref{cap:xi_dataMEM}.
The resulting $\nu$ was in much better agreement with the GL model prediction.  Likewise, $\beta$ was much closer
to $1/2$ than for the FT method.  Averaged over $\Omega$, $\nu\simeq0.46$ and $\beta\simeq0.63$ from the MEM.  
Figure \ref{cap:nu_Omega} summarizes the
behavior of the exponents for both the MEM and the FT method as a function of $\Omega$.  The MEM clearly gave
the closest result in agreement with the prediction $\nu=1/2$ and also $\beta=1/2$.  However, the FT method
yielded the best agreement with the prediction $\nu+\beta=1$, with the MEM not much further off.  Considering the results of Sect.~\ref{sub:SimSH},
the MEM is more reliable.  Thus we conclude that the experimental length scale is in agreement with the prediction from the GL model.

\section{\label{sub:Scaling_Collapse}Scaling of $\delta S\left(k\right)$
for Experimental Data and Analytic Functions}

\begin{figure}
\begin{center}\includegraphics[width=3.33in, keepaspectratio]{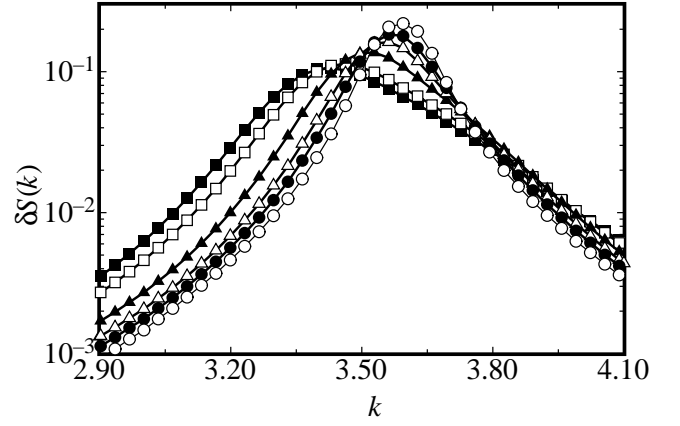}\end{center}
\caption{\label{cap:Normalized_SF} The normalized SFs for $\Omega=17.7$ computed with the MEM. 
Open circles: $\varepsilon = 0.017$.  Solid circles: $\varepsilon = 0.025$.  
Open triangles: $\varepsilon = 0.033$.  Solid triangles: $\varepsilon = 0.05$.  
Open squares: $\varepsilon = 0.08$.  Solid squares: $\varepsilon = 0.10$.}
\end{figure}

The scaling analogy to critical phenomena extends beyond the power-law dependence of $\xi$ and $B$ on $\varepsilon$. In order to more deeply probe this scaling,
we re-scaled the SFs at different $\varepsilon$ in an attempt to collapsed them all
onto a single curve. We followed the procedure
of Ref.~\cite{XLG97}. First we normalized the structure factor so that $2\pi$$\int_{0}^{\infty}\delta S\left(k\right)kdk=1$.
In applying this normalization to the experimental data, we evaluated the integral over the range
$k_0-3/\xi\leq k \leq k_0+3/\xi$.  This included most of the total power present in the experimental data.
Figure \ref{cap:Normalized_SF} shows the MEM results
for several $\varepsilon$ values and demonstrates that normalization alone
is not sufficient to collapse the data onto a unique curve. It is also necessary to re-scale
the SF on both the abscissa and ordinate axes so that $x\equiv\left(k-k_{0}\right)\xi$
and $\delta\tilde{S}\left(x\right)\equiv\left(k/\xi\right)\delta S\left(x\right)$.

Applying this variable transformation to the analytic forms in Eqs.~\ref{eq:SHSF}-\ref{eq:LorentzianSF}
provides some insight into the effect of rescaling. In the critical-point limit, of $\xi\rightarrow\infty$ and
evaluating the integral normalization over the range $0\leq k\leq\infty$, all parameters cancel, 
leaving only a function of $x$. This yields

\begin{equation}
\tilde{S}\left(x\right)=\frac{1}{2\pi^{2}\left(1+x^{2}\right)}\label{eq:SH_rescaled}\end{equation}

\noindent in the case of the SH form,

\begin{equation}
\tilde{S}\left(x\right)=\frac{1}{\pi^{2}\left(1+x^{2}\right)^{2}}\label{eq:squaredSH_rescaled}\end{equation}

\noindent in the case of the squared SH form, and

\begin{equation}
\tilde{S}\left(x\right)=\frac{\exp\left(-x^{2}\right)}{2\pi^{3/2}}\label{eq:Gaussian_rescaled}\end{equation}

\noindent in the case of the Gaussian form. We refrain from listing the rescaled
Lorentzian because the zeroth moment diverges, making it impossible
to normalize without introducing a cutoff.

\begin{figure}
\begin{center}\includegraphics[width=3.33in, keepaspectratio]{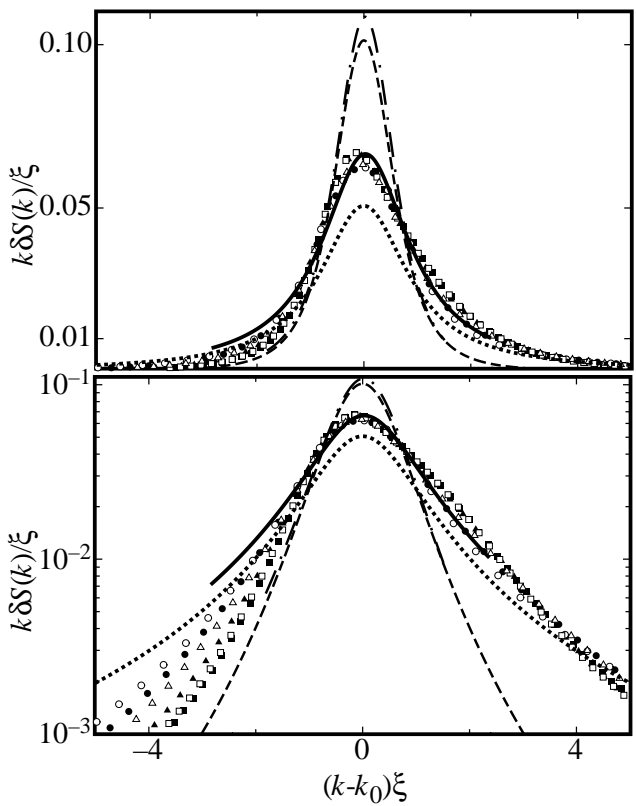}\end{center}
\caption{\label{cap:Collapsed_SF}Scaled SFs for $\Omega=17.7$ computed with the MEM. 
Top figure: linear-linear plot. Bottom figure: log-linear plot. 
The symbols are as in Fig.~\ref{cap:Normalized_SF}. Dotted line: rescaled SH SF Eq.~\ref{eq:SH_rescaled}.
Dashed line: rescaled
squared SH SF Eq.~\ref{eq:squaredSH_rescaled}. Solid line: numerically rescaled SH SF using finite limits for its normalization.
Dashed-dotted line: numerically rescaled squared SH SF using finite limits for its normalization.}
\end{figure}

Figure \ref{cap:Collapsed_SF} shows the re-scaled experimental SFs from the MEM on a linear (top figure) and on a logarithmic (bottom figure) scale.  Although not shown, we found comparable results for both the normalized and rescaled SFs computed with
the FT method.  The figure also includes the rescaled SH and squared SH SFs given by Eqs.~\ref{eq:SH_rescaled} and \ref{eq:squaredSH_rescaled}.
We omit the Gaussian form in Eq.~\ref{eq:Gaussian_rescaled} in order to maximize the clarity of the figure.  It is the 
poorest empirical representation out of Eqs.~\ref{eq:SH_rescaled}-\ref{eq:Gaussian_rescaled}, and the least
physically justifiable.
Note that the curves of Eqs.~\ref{eq:SH_rescaled}
and \ref{eq:squaredSH_rescaled} are not fits to
the experimental data; there are no free parameters for fitting. None of them provide a particularly good scaled representation of the data near the peak. The experimental results fall somewhere between the SH (dotted line) and the squared SH (dashed line) form.

The disagreement between the scaled data and the scaled functions is primarily due to the normalization of the experimental data, which is over a finite range
unlike the normalization in Eqs.~\ref{eq:SH_rescaled} and \ref{eq:squaredSH_rescaled} which is over all $k$.
Also shown are collapsed forms of Eqs.~\ref{eq:SHSF} and \ref{eq:squaredSHSF} computed numerically using the 
normalization over the same finite range as in the experimental SFs.  The numerically-computed forms were 
evaluated at finite $\varepsilon$, instead of taking the limit of $\xi\rightarrow\infty$, as was done in 
the analytic cases of Eqs.~\ref{eq:SH_rescaled}-\ref{eq:Gaussian_rescaled}. 
As a result, $\xi$ and $k_0$ remained as constants in the collapsed forms.  We used the values of $\xi$ and $k_0$ from the $\varepsilon=0.10$ data point.
This choice barely affected the numerically collapsed curves
because $k_0\xi$ was relatively large for all values of $\varepsilon$ shown.
These curves provide insight into the agreement of
the shape of Eqs.~\ref{eq:SHSF} and \ref{eq:squaredSHSF} with the experimental data.  
The numerically computed SH SF agreed quite well with the collapsed experimental data, while the squared SH SF did not.

\section{Summary and Conclusion}

In this paper we carefully examined two methods, the FT method  and the MEM \cite{NR}, for determining critical parameters describing the spatio-temporal chaos state in rotating Rayleigh-B\'enard convection known as domain chaos. We found that a correlation length $\xi$, equal to the inverse half-width at half-height of the structure factor $S(k)$, can be determined reliably by fitting one of several trial functions to the experimental data. We examined a Lorentzian, Swift-Hohenberg, squared Swift-Hohenberg, and Gaussian form. We prefer this fitting method over computing numerical moments from the data (as had been done in the past) because we were able to show analytically that the length $\bar \xi = 1/\sigma$ ($\sigma^2$ is the variance of the data) will be proportional to $\xi$ only when $S(k)$ decreases sufficiently fast at large $k$.

For a technique of estimating the SF, we found that the MEM was much superior to the classic
FT method.  There is a severe dependence on image size in the case of the FT method that is quite apparent from our
analysis of various sized images from the simulation using the SH domain-chaos model.  Since the experimental images
are relatively small, the MEM is required in order to accurately determine the SF from the patterns.

We analyzed new experimental shadowgraph images of domain chaos for a sample of aspect ratio $\Gamma = 61.5$ over the range $15 \stackrel {<}{_\sim} \Omega \stackrel {<}{_\sim} 22$. Using the above four functional forms as fitting functions, we obtained results for $\xi$ that depended only slightly on the function used, but that all had the same dependence on the distance $\varepsilon$ from the onset of convection. In the case of the FT method, we found $\xi \sim \varepsilon^{-\nu}$ with $\nu \simeq 0.25$, in agreement with previous measurements for a sample with $\Gamma = 40$ but in disagreement with expectations based on the weakly-nonlinear amplitude model. Fortunately the MEM was able to
overcome the image-size problem of the FT method and yielded $\nu\simeq0.46$, roughly in agreement with the theoretical models. We also determined the maximum height $B$ of $S(k)$ from fits of the functions to the data and found that $B \sim \varepsilon^{\beta}$ with $\beta \simeq 0.73$ for the FT method, in disagreement with the value $\beta = 1/2$ suggested by the amplitude model. However, we note that the FT method yielded a total power proportional to $\varepsilon^{\nu+\beta}$ with $\nu + \beta \simeq 0.98$, that agrees with the theoretical expectation that $\nu + \beta = 1$.  The MEM yielded $\beta\simeq0.63$, somewhat closer to the prediction of $\beta=1/2$ than the FT method.
		 
We also showed that it is possible to present the structure factor in a scaled form that largely collapses the data onto a unique curve.  This was the case for both the FT method and the MEM, even though the scaling exponents of the FT method results
did not agree with the prediction from theory.

In summary, our new experimental data and analysis yielded results that differ from the earlier result that $\nu \simeq 0.25$ in disagreement with theory.  We have shown that this difference is due to the limitations of the FT analysis method.  The MEM is
capable of extracting scaling exponents from the data that are in agreement with the theoretical prediction.  In addition to the exponent $\nu$, we also examined a scaling parameter $\beta$ that describes the height of the structure factor, and found it to also agree with predictions when the MEM results were used.

\section{Acknowledgment}

We are grateful for numerous stimulating conversations with a number of people, including especially E. Bodenschatz, M.C. Cross, P.C. Hohenberg, M. Paul, W. Pesch, and J.D. Scheel. This work was supported by the US National Science Foundation through Grant DMR02-43336.


\begin{thebibliography}{10}
\bibitem{FNsize} We distinguish between finite-sample-size effects due to the lateral boundaries that were discussed in Ref.~\cite{CLM01}, and finite image-size effects that would prevail even if a small image were extracted from a central region of a much larger sample.
\end{thebibliography}
\end{document}